\documentclass[a4paper,10pt]{article}
\usepackage{amssymb}
\usepackage{amsmath}
\usepackage{epsfig}
\usepackage{subfig}
\usepackage{caption}
\usepackage{graphicx}
\usepackage{tikz}
\usepackage{latexsym}
\usepackage{rotating}
\usepackage{titlesec}

\usetikzlibrary{matrix}
\newtheorem{thm}{Theorem}[section]

\newtheorem{rem}[thm]{Remark}

\usepackage{relsize}

\title{Method of ${\cal M}_{n}$-Extension via Frobenius Companion Matrices}

\author{Metin G\"{u}rses \thanks{gurses@fen.bilkent.edu.tr}\\
{\small Department of Mathematics, Faculty of Science}\\
{\small Bilkent University, 06800 Ankara - Turkey}\\
Asl{\i} Pekcan \thanks{Corresponding Author: aslipekcan@hacettepe.edu.tr} \\
{\small Department of Mathematics, Faculty of Science} \\
{\small Hacettepe University, 06800 Ankara - Turkey}
}

\date{\nonumber}
\begin{document}
\maketitle
\date{\nonumber}

\begin{abstract}
Frobenius companion matrices arise when we write an $n$-th order linear ordinary differential equation as a system of first order differential equations. These matrices and their transpose have very nice properties. By using the powers of these matrices we form a closed algebra under the matrix multiplication. Structure constants of this commuting algebra are the components of companion matrix. We use these matrices in our method of ${\cal M}_{n}$-extension of scalar integrable equations to produce new systems of integrable equations with recursion operators.\\

\noindent \textbf{Keywords}: ${\cal M}_{n}$-extension, Companion matrices, Structure coefficients, Coupled systems, Integrability.
\end{abstract}


\section{Introduction}

Recently, in \cite{gur-pek1}, \cite{gur-pek2} we introduced a method called ${\cal M}_{n}$-extension of scalar integrable equations to obtain new systems of integrable equations. The basic idea in this method is to express the scalar dynamical variable as a matrix function. Then writing this matrix function in a closed algebra we derive the system of integrable equations and their recursion operators in terms of structure constants of this commuting algebra. There are also attempts \cite{ma0}-\cite{ma2} to use similar approaches to obtain integrable systems of equations. Frobenius companion matrices and their powers form such an algebra. We will use these matrices to produce systems of Korteweg-de Vries (KdV), modified KdV (MKdV), Sawada-Kotera (SK), Kaup-Kupershmidt (KK), and nonlinear Schr\"{o}dinger (NLS) equations.

Companion matrices  have several applications  such as canonical forms, matrix inequalities, interpolations, differential equations, and difference equations \cite{Duf}-\cite{comp}. For instance, if we wish to write $n$-th order linear ordinary differential equation as a system of first order ordinary differential equations (ODEs) we end up with such a matrix.

Let an $n$-th order linear ODE be given by
\begin{equation}
y^{(n)}=c_{n-1}\, y^{(n-1)}+\cdots+c_{1}\, y^{(1)}+c_{0} y,
\end{equation}
where $c_{i}$'s are constants ($i=0,1,\cdots, n-1$). Then letting $z(t)=(y, y^{(1)}, \cdots , y^{(n-1)})$ we obtain the following first order system of linear ODEs
\begin{equation}
\dot{z}=M z,
\end{equation}
where $M$ and its transpose are called the companion matrices. It has the following form:
{\small \begin{equation}
M=\left( \begin{array}{ccccccc}
0&1&0&0&\ldots&0&0 \\
0&0&1&0&\ldots&0&0\\
0&0&0&1&\ldots&0&0\\
\vdots&\vdots&\vdots&\ddots&\vdots&\vdots&\vdots\\
0&0&0&0&0&0&1\\
c_{0}&c_{1}&c_{2}&c_{3}&\ldots&c_{n-2}&c_{n-1}
\end{array} \right).
\end{equation}}

These matrices have interesting properties. We shall make use of these properties in our ${\cal M}_{n}$-extension method.

Let $A$ be an $n \times n$ such a companion matrix.
By Cayley-Hamilton theorem \cite{comp} every square matrix $A_{n\times n}$ satisfies its own characteristic equation, i.e.,
if the characteristic equation of the matrix $A_{n\times n}$ is
\begin{equation}
\mathrm{det}(A-\lambda I)=\lambda^n-a_{n-1}\lambda^{n-1}-\ldots-a_1\lambda-a_0=0
\end{equation}
then
\begin{equation}\label{A^N}
A^n=a_{n-1}A^{n-1}+\cdots+a_1 A+a_0.
\end{equation}

We can construct from the powers of $A$ a commutative and associative algebra with the basis
\begin{equation*}
 \Sigma=\{ \Sigma_{0}=I, \Sigma_{1}=A, \Sigma_{2}=A^2, \cdots, \Sigma_{n-1}=A^{n-1}\}
 \end{equation*}
satisfying the following product rule:
\begin{equation}\label{cond1}
\Sigma_i\cdot\Sigma_j=f_{ij}^k\Sigma_k,~~~i,j=0,1,\cdots n-1,
\end{equation}
where we use summation convention for the repeated indices and $f_{ij}^k$ are the structure constants of the algebra which are symmetrical with respect to the indices $i$ and $j$, i.e., $f_{ij}^k=f_{ji}^k$. Due to associativity of the matrix product the structure constants $f_{ij}^k$ satisfy
\begin{equation}\label{assoc}
f^{k}_{ij}\, f^{r}_{k \ell}=f^{k}_{\ell i}\, f^{r}_{k j},
\end{equation}
where $i,j,k,r,\ell=0,1,2,\cdots, n-1$. The structure constants and the matrices $\Sigma_{i}$ have also a nice correspondence
\begin{equation}\label{cond2}
 (\Sigma_k)_b^a=f_{kb}^a,~~~k,a,b=0,1,\cdots, n-1.
\end{equation}

In the sequel, all our companion matrices will be of the form
{\small\begin{equation}\label{frob1}
\Sigma_1=A=\left( \begin{array}{cccccc}
0&0&0&\ldots&0& a_{0}  \\
1&0&0&\ldots&0&a_{1}\\
0&1&0&\ldots&0&a_{2}\\
0&0&1&\ldots&0&a_{3}\\
\vdots&\vdots&\vdots&\ddots&\vdots&\vdots\\
0&0&0&\ldots&1&a_{n-1}
\end{array} \right).
\end{equation}}

\begin{rem}
 If we start with an arbitrary $n \times n$ matrix $A$ and form the closed algebra $\{\Sigma_{0}=I, \Sigma_{1}=A, \Sigma_{2}=A^2, \cdots, \Sigma_{n-1}=A^{n-1}\}$
and calculate the structure constant $f^{k}_{ij}$ from (\ref{cond1}) and impose the condition (\ref{cond2}) then we end up with Frobenius companion matrix (\ref{frob1}).
\end{rem}

\noindent
For illustration how to use the companion matrices in our method we shall give the Burgers' equation,
\begin{equation}
u_{t}=u_{xx}+2 u u_{x}.
\end{equation}

\vspace{0.2cm}
\noindent
(1) First we let $u \to U=\Sigma_{i}\,u^{i}=u^{0} I+\Sigma_{1} u^{1}+\cdots+\Sigma_{n-1}\,u^{n-1}$, where $I$ is the $n \times n$ unit matrix and $u^{i}$'s ($i=0,1,\cdots, n-1$) are dynamical variables depending on $x$ and $t$. Later we shall take $u^{0}=u$.

\vspace{0.2cm}
\noindent
(2) By letting $u \to U$, the Burgers' equation becomes
\begin{equation}
U_{t}=U_{xx}+2 U U_{x}.
\end{equation}
Here the order in $U U_{x}$ is not important because the algebra is commutative.
To find the evolution equations for $u^{i}$'s ($i=0,1,\cdots,n-1$), using the product rule (\ref{cond1}) of $\Sigma_{i}$'s ($i=0,1,\cdots, n-1$), we write
the nonlinear terms in the scalar equations in a more simple form. In the case of the Burgers' equation, $U U_{x}$ is such a term which can be written as
\begin{equation}
U U_{x}=(\Sigma_{i} u^{i}) \, ( \Sigma_{j} u_x^{j} )= \left(f^{k}_{ij}\, u^{i} u_x^{j} \right)\,\Sigma_{k}.
\end{equation}
Then we produce the following system of evolution equations from the Burgers' equation:
\begin{equation}\label{sys}
u^{i}_{t}=u^{i}_{xx}+2 f^{i}_{k \ell}\, u^{k} u_x^{\ell},~~~(i=0,1,\cdots,n-1).
\end{equation}

\vspace{0.3cm}
\noindent
(3) The above Burgers' system admits a recursion operator which can be obtained from the recursion operator of the Burgers' equation
\begin{equation}
R_{Burg}=D +u +u_{x} D^{-1}.
\end{equation}
Then letting $u \to U$ we get
\begin{equation}
{\cal{R}}=I D +U +U_{x} D^{-1}=I D+u^{k} \Sigma_{k}+u^{k}_{x}\Sigma_{k}\, D^{-1}.
\end{equation}
Using  (\ref{cond2}) we get the components of the recursion operator of the system of equations (\ref{sys})
\begin{equation}
{\cal{R}}^{a}_{b}=\delta^{a}_{b}D+f^{a}_{bc} u^{c}+f^{a}_{bc}u^{c}_{x} D ^{-1},
\end{equation}
which implies
\begin{equation}
u^{a}_{t}={\cal{R}}^{a}_{b}\, u^{b}_{x},  ~~~(a=0,1,\cdots,n-1).
\end{equation}

In the next sections we shall give more examples including KdV, MKdV, SK, and KK equations, and NLS system. Note that one can obtain new integrable nonlocal unshifted and shifted
equations from multi-field extensions of such integrable equations. We considered nonlocal reductions of the ${\cal M}_{2}$-extension of KdV equation in \cite{gur-pek1}. We also studied nonlocal reductions of the ${\cal M}_{2}$-extension of SK and KK equations in \cite{gur-pek2}. We shall consider different examples of nonlocal reductions of ${\cal M}_{n}$-extensions obtained by our new approach in more detail in a forthcoming publication.

\section{Structure constants of the algebra for $n=2,3,4$}

\subsection{n=2}

 For $n=2$, i.e., for $2\times 2$ matrix $A$, from Cayley-Hamilton theorem we  have
\begin{equation}\label{CH-N=2}
A^2=(\mathrm{tr}\,A)A-(\mathrm{det}\,A)I,
\end{equation}
where $I$ is the $2\times 2$ identity matrix.

\noindent Take $\Sigma_0=I$ and $\Sigma_1=A$. Let also $\mathrm{tr}\,A=\alpha$, $\mathrm{det}\,A=-\beta$. Therefore the relation (\ref{CH-N=2}) becomes
\begin{equation}
A^2=\alpha A+\beta I.
\end{equation}
The condition (\ref{cond1}) gives the structure coefficients as
\begin{equation}
f_{00}^0=1,\,\, f_{01}^0=0,\,\, f_{11}^0=\beta,\,\, f_{01}^1=1,\,\, f_{11}^1=\alpha.
\end{equation}
For consistency, we check the condition (\ref{cond2}). We obtain that
{\small\begin{equation}\label{Sigma_1n=2}
\Sigma_1=A=\left( \begin{array}{cc}
0&\beta  \\
1&\alpha
\end{array} \right).
\end{equation}}

\subsection{n=3}

For $n=3$, from Cayley-Hamilton theorem we have
\begin{equation}\label{CH-N=3}
A^3=(\mathrm{tr}\,A)A^2-\frac{1}{2}[(\mathrm{tr}\,A)^2-\mathrm{tr}\,(A^2)]A+(\mathrm{det}\,A)I,
\end{equation}
where $A$ is $3\times 3$ matrix and $I$ is the $3\times 3$ identity matrix.

\noindent Choose $\Sigma_0=I$, $\Sigma_1=A$, and $\Sigma_2=A^2$. Let $\mathrm{tr}\,A=\alpha$, $\mathrm{det}\,A=-\beta$, and
$\frac{1}{2}[(\mathrm{tr}\,A)^2-\mathrm{tr}\,(A^2)]=\gamma$. Hence the relation (\ref{CH-N=3}) turns to be
\begin{equation}
A^3=\alpha A^2-\gamma A-\beta I.
\end{equation}
From (\ref{cond1}) we obtain
{\small
\begin{align}
&f_{00}^0=1,\,\, f_{01}^0=f_{02}^0=f_{01}^0=f_{11}^0=f_{02}^0,\,\, f_{12}^0=-\beta,\,\, f_{22}^0=-\alpha\beta,\\
&f_{00}^1=f_{02}^1=f_{11}^1=0,\,\, f_{01}^1=1,\,\, f_{12}^1=-\gamma,\,\, f_{22}^1=-\alpha\gamma-\beta,\\
&f_{00}^2=f_{01}^2=0,\,\, f_{02}^2=1,\,\, f_{11}^2=1,\,\, f_{12}^2=\alpha, \,\, f_{22}^2=\alpha^2-\gamma.
\end{align}}
Consider now the condition (\ref{cond2}).

\noindent For $k=0$, we have
\begin{equation}
(\Sigma_0)_b^a=f_{0b}^a=f_{b0}^a=\delta_b^a.
\end{equation}

\noindent For $k=1$, we have $(\Sigma_1)_b^a=f_{b1}^a=f_{1b}^a$. Explicitly,
\begin{align}
&(\Sigma_1)_0^0=f_{01}^0=0,\,\, (\Sigma_1)_1^0=f_{11}^0=0,\,\, (\Sigma_1)_2^0=f_{21}^0=-\beta,\\
&(\Sigma_1)_0^1=f_{01}^1=1,\,\, (\Sigma_1)_1^1=f_{11}^1=0,\,\, (\Sigma_1)_2^1=f_{21}^1=-\gamma,\\
&(\Sigma_1)_0^2=f_{01}^2=0,\,\,  (\Sigma_1)_1^2=f_{11}^2=1,\,\, (\Sigma_1)_2^2=f_{21}^2=\alpha.
\end{align}
Hence,
{\small\begin{equation}\label{Sigma_1n=3}
\Sigma_1=A=\left( \begin{array}{ccc}
0&0&-\beta  \\
1&0&-\gamma\\
0&1&\alpha
\end{array} \right).
\end{equation}}
Here, it is clear that $\mathrm{tr}\,A=\alpha$ and $\mathrm{det}\,A=-\beta$.

\noindent For $k=2$, we have $(\Sigma_2)_b^a=f_{b2}^a=f_{2b}^a$ yielding
{\small
\begin{align}
&(\Sigma_2)_0^0=f_{02}^0=0,\,\, (\Sigma_2)_1^0=f_{12}^0=-\beta,\,\, (\Sigma_2)_2^0=f_{22}^0=-\alpha\beta,\\
&(\Sigma_2)_0^1=f_{02}^1=0,\,\, (\Sigma_2)_1^1=f_{12}^1=-\gamma,\,\, (\Sigma_2)_2^1=f_{22}^1=-\alpha\gamma-\beta,\\
&(\Sigma_2)_0^2=f_{02}^2=1,\,\, (\Sigma_2)_1^2=f_{12}^2=\alpha,\,\, (\Sigma_2)_2^2=f_{22}^2=\alpha^2-\gamma.
\end{align}}
Hence,
{\small\begin{equation}\label{Sigma_2n=3}
\Sigma_2=A^2=\left( \begin{array}{ccc}
0&-\beta&-\alpha\beta  \\
0&-\gamma&-\alpha\gamma-\beta\\
1&\alpha&\alpha^2-\gamma
\end{array} \right).
\end{equation}}

\subsection{n=4}

For $n=4$, Cayley-Hamilton theorem gives
\begin{align}\label{CH-N=4}\displaystyle
A^4=(\mathrm{tr}\,A)A^3-\frac{1}{2}[(\mathrm{tr}\,A)^2-\mathrm{tr}\,(A^2)]A^2+\frac{1}{6}[(\mathrm{tr}\,A)^3-3\mathrm{tr}\,(A^2)\mathrm{tr}\,A+2\mathrm{tr}\,(A^3)]A-(\mathrm{det}\,A)I,
\end{align}
where $A$ is $4\times 4$ matrix and $I$ is the $4\times 4$ identity matrix.

\noindent Take $\Sigma_0=I$, $\Sigma_1=A$, $\Sigma_2=A^2$, and $\Sigma_3=A^3$. Let $\mathrm{tr}\,A=\alpha$, $\mathrm{det}\,A=-\beta$, $\frac{1}{2}[(\mathrm{tr}\,A)^2-\mathrm{tr}\,(A^2)]=\gamma$,
and $\frac{1}{6}[(\mathrm{tr}\,A)^3-3\mathrm{tr}\,(A^2)\mathrm{tr}\,A+2\mathrm{tr}\,(A^3)]=\mu$. Hence the relation (\ref{CH-N=4}) becomes
\begin{equation}
A^4=\alpha A^3-\gamma A^2+\mu A+\beta I.
\end{equation}
From (\ref{cond1}) we obtain the structure coefficients as
{\small \begin{align}
&f_{00}^0=1,\,\, f_{01}^0=f_{02}^0=f_{03}^0=f_{11}^0=f_{12}^0=0,\,\, f_{13}^0=f_{22}^0=\beta, f_{23}^0=\alpha\beta,\,\, f_{33}^0=(\alpha^2-\gamma)\beta,\\
&f_{00}^1=f_{02}^1=f_{03}^1=f_{11}^1=f_{12}^1=0,\,\, f_{01}^1=1,\,\, f_{13}^1=f_{22}^1=\mu,\,\, f_{23}^1=\alpha\mu+\beta,\nonumber\\
&f_{33}^1=\alpha^2\mu-\gamma\mu+\alpha\beta,\\
&f_{00}^2=f_{01}^2=f_{03}^2=f_{12}^2=0,\,\, f_{02}^2=f_{11}^2=1,\,\, f_{13}^2=f_{22}^2=-\gamma,\,\, f_{23}^2=-\alpha\gamma+\mu,\nonumber\\ &f_{33}^2=-\alpha^2\gamma+\gamma^2+\alpha\mu+\beta,\\
&f_{00}^3=f_{01}^3=f_{02}^3=f_{11}^3=0,\,\, f_{03}^3=f_{12}^3=1,f_{13}^3=f_{22}^3=\alpha,\,\, f_{23}^3=\alpha^2-\gamma,\nonumber\\
&f_{33}^3=\alpha^3-2\alpha\gamma+\mu.
\end{align}}
Consider now the condition (\ref{cond2}).

\noindent For $k=0$, we have
\begin{equation}
(\Sigma_0)_b^a=f_{0b}^a=f_{b0}^a=\delta_b^a.
\end{equation}

\noindent For $k=1$, we have $(\Sigma_1)_b^a=f_{b1}^a=f_{1b}^a$ giving
{\small\begin{align}
&(\Sigma_1)_0^0=f_{01}^0=0,\,\, (\Sigma_1)_1^0=f_{11}^0=0,\,\, (\Sigma_1)_2^0=f_{21}^0=0,\,\, (\Sigma_1)_3^0=f_{31}^0=\beta,\\
&(\Sigma_1)_0^1=f_{01}^1=1,\,\, (\Sigma_1)_1^1=f_{11}^1=0,\,\, (\Sigma_1)_2^1=f_{21}^1=0,\,\, (\Sigma_1)_3^1=f_{31}^1=\mu,\\
&(\Sigma_1)_0^2=f_{01}^2=0,\,\, (\Sigma_1)_1^2=f_{11}^2=1,\,\, (\Sigma_1)_2^2=f_{21}^2=0,\,\, (\Sigma_1)_3^2=f_{31}^2=-\gamma,\\
&(\Sigma_1)_0^3=f_{01}^3=0,\,\, (\Sigma_1)_1^3=f_{11}^3=0,\,\, (\Sigma_1)_2^3=f_{21}^3=1,\,\, (\Sigma_1)_3^3=f_{31}^3=\alpha.
\end{align}}
Therefore,
{\small\begin{equation}\label{Sigma_1n=4}
\Sigma_1=A=\left( \begin{array}{cccc}
0&0&0&\beta  \\
1&0&0&\mu\\
0&1&0&-\gamma\\
0&0&1&\alpha
\end{array} \right).
\end{equation}}
It is clear that $\mathrm{tr}\,A=\alpha$ and $\mathrm{det}\,A=-\beta$.

\noindent For $k=2$, we have $(\Sigma_2)_b^a=f_{b2}^a=f_{2b}^a$ yielding
{\small\begin{align}
&(\Sigma_2)_0^0=f_{02}^0=0,\,\, (\Sigma_2)_1^0=f_{12}^0=0,\,\, (\Sigma_2)_2^0=f_{22}^0=\beta,\,\, (\Sigma_2)_3^0=f_{32}^0=\alpha\beta,\\
&(\Sigma_2)_0^1=f_{02}^1=0,\,\, (\Sigma_2)_1^1=f_{12}^1=0,\,\, (\Sigma_2)_2^1=f_{22}^1=\mu,\,\, (\Sigma_2)_3^1=f_{32}^1=\alpha\mu+\beta,\\
&(\Sigma_2)_0^2=f_{02}^2=1,\,\, (\Sigma_2)_1^2=f_{12}^2=0,\,\, (\Sigma_2)_2^2=f_{22}^2=-\gamma,\,\, (\Sigma_2)_3^2=f_{32}^2=-\alpha\gamma+\mu,\\
&(\Sigma_2)_0^3=f_{02}^3=0,\,\, (\Sigma_2)_1^3=f_{12}^3=1,\,\, (\Sigma_2)_2^3=f_{22}^3=\alpha,\,\, (\Sigma_2)_3^3=f_{32}^3=\alpha^2-\gamma.
\end{align}}
Hence,
{\small\begin{equation}
\Sigma_2=A^2=\left( \begin{array}{cccc}
0&0&\beta&\alpha\beta  \\
0&0&\mu&\alpha\mu+\beta\\
1&0&-\gamma&-\alpha\gamma+\mu\\
0&1&\alpha&\alpha^2-\gamma
\end{array} \right).
\end{equation}}

\noindent Finally, for $k=3$, we have $(\Sigma_3)_b^a=f_{b3}^a=f_{3b}^a$ giving
{\small\begin{align}
&(\Sigma_3)_0^0=f_{03}^0=0,\,\, (\Sigma_3)_1^0=f_{13}^0=\beta,\,\, (\Sigma_3)_2^0=f_{23}^0=\alpha\beta,\,\, (\Sigma_3)_3^0=f_{33}^0=(\alpha^2-\gamma)\beta,\\
&(\Sigma_3)_0^1=f_{03}^1=0,\,\, (\Sigma_3)_1^1=f_{13}^1=\mu,\,\, (\Sigma_3)_2^1=f_{23}^1=\alpha\mu+\beta,\,\, (\Sigma_3)_3^1=f_{33}^1=\alpha^2\mu-\gamma\mu+\alpha\beta,\\
&(\Sigma_3)_0^2=f_{03}^2=0,\,\, (\Sigma_3)_1^2=f_{13}^2=-\gamma,\,\, (\Sigma_3)_2^2=f_{23}^2=-\alpha\gamma+\mu,\,\, (\Sigma_3)_3^2=f_{33}^2=\gamma^2-\alpha^2\gamma+\alpha\mu+\beta,\\
&(\Sigma_3)_0^3=f_{03}^3=1,\,\, (\Sigma_3)_1^3=f_{13}^3=\alpha,\,\, (\Sigma_3)_2^3=f_{23}^3=\alpha^2-\gamma,\,\, (\Sigma_3)_3^3=f_{33}^3=\alpha^3-2\alpha\gamma+\mu.
\end{align}}
Hence,
{\small \begin{equation}
\Sigma_3=A^3=\left( \begin{array}{cccc}
0&\beta&\alpha\beta&(\alpha^2-\gamma)\beta  \\
0&\mu&\alpha\mu+\beta&\alpha^2\mu-\gamma\mu+\alpha\beta\\
0&-\gamma&-\alpha\gamma+\mu&\gamma^2-\alpha^2\gamma+\alpha\mu+\beta\\
1&\alpha&\alpha^2-\gamma&\alpha^3-2\alpha\gamma+\mu
\end{array} \right).
\end{equation}}

We assume that components of the companion matrix ($\alpha, \beta, \gamma, \mu, \cdots$) are all complex numbers in general.

\section{KdV system}

The KdV equation and  its recursion operator are given as
\begin{eqnarray}
&&u_{t}=u_{xxx}+6 u u_{x}, \label{KdV}\\
&&R_{KdV}=D^2+4u+2u_{x}\, D^{-1}. \label{recKdV}
\end{eqnarray}

The ${\cal M}_{n}$-extension of the KdV equation is obtained by letting $u \to U= u^{k}\, \Sigma_{k}=u^{0} I+u^{\alpha}\, \Sigma_{\alpha}$, ($\alpha=1,2 \cdots n-1$). We have
the following system of equations:
\begin{equation}\label{M_NKdV}
u^{i}_{t}=u^{i}_{xxx}+6 f^{i}_{jk}\, u^{j} u^{k}_{x},~~(i=0,1,2, \cdots, n-1),
\end{equation}
and the components of the recursion operator are
\begin{equation}
{\cal{R}}^{a}_{b}=\delta^{a}_{b} D^2+4 f^{a}_{bc} u^c+2 f^{a}_{bc} u^{c}_{x}\, D^{-1},
\end{equation}
giving
\begin{equation}
u^{a}_{t}={\cal{R}}^{a}_{b}\, u^{b}_{x},  ~~~(a=0,1,\cdots,n-1).
\end{equation}

\noindent \textbf{Case n=2.}
To obtain ${\cal M}_{2}$-extension of the KdV equation \cite{gur-pek2}-\cite{ma2} let $u\rightarrow U=u\Sigma_0+v\Sigma_1$ where $\Sigma_0=I$ and $\Sigma_1$ is given by (\ref{Sigma_1n=2}), i.e., satisfying $\Sigma_1^2=\alpha \Sigma_1+\beta I$ for
$\alpha=\mathrm{tr}\,\Sigma_1$, $\beta=-\mathrm{det}\,\Sigma_1$. We have $U_t=U_{xxx}+6UU_x$ which is explicitly corresponds to the following coupled KdV system
\begin{align}
&u_t=u_{xxx}+6uu_x+6\beta vv_x,\\
&v_t=v_{xxx}+6(uv)_x+6\alpha vv_x.
\end{align}
Recursion operator of this system is
\begin{equation}
{\cal{R}}=\left(\begin{array}{ll}
        R_{KdV} &\quad \quad \quad \beta  (4 v+2 v_{x}\, D^{-1})\cr
        4 v+2 v_{x}\, D^{-1} &\quad \quad R_{KdV}+ \alpha  (4 v+2 v_{x}\, D^{-1})\cr
        \end{array} \right).
\end{equation}

\vspace{0.5cm}

\noindent \textbf{Case n=3.} Consider the ${\cal M}_{3}$-extension of KdV equation (\ref{KdV}). Let $u\rightarrow U=uI+v\Sigma_1+w\Sigma_2$ where $\Sigma_1$ is given by (\ref{Sigma_1n=3}). We have $U_t=U_{xxx}+6UU_x$ which is explicitly corresponds to the following coupled KdV system of three equations:
\begin{align}
&u_t=u_{xxx}+6uu_x-6\beta (vw)_x-6\alpha\beta ww_x,\\
&v_t=v_{xxx}+6(uv)_x-6\gamma (vw)_x-6(\alpha\gamma+\beta)ww_x,\\
&w_t=w_{xxx}+6(uw)_x+6\alpha (vw)_x+6(\alpha^2-\gamma)ww_x.
\end{align}
Recursion operator of the above system is
\begin{equation}
{\cal{R}}=\left(\begin{array}{lll}
        R_{KdV} &\quad  -\beta  W_2 &\quad -\beta W_1-\beta\alpha W_2  \cr
        W_1 &\quad R_{KdV}- \gamma W_2 &\quad -\gamma W_1-(\alpha\gamma+\beta) W_2\cr
        W_2&\quad W_1+\alpha W_2 &\quad R_{KdV}+\alpha W_1+(\alpha^2-\gamma) W_2
        \end{array} \right),
\end{equation}
where $W_1=4 v+2 v_{x}\, D^{-1}$ and $W_2=4 w+2 w_{x}\, D^{-1}$.

Systems of KdV equation and their recursion operators have been studied by several groups \cite{gur-kar1}-\cite{Svin1}. Our method of ${\cal M}_{n}$-extension produces some of these systems easily and explicitly with the structure constants for each $n$.

\section{MKdV system}

The MKdV equation and its recursion operator are
\begin{align}
&u_t=u_{xxx}+6 u^2u_x,\label{MKdV}\\
&R_{MKdV}=D^2+4 u^2+4 u_{x} D^{-1}u.\label{recMKdV}
\end{align}

The ${\cal M}_{n}$-extension of the MKdV equation obtained by taking $u \to U= u^{k}\, \Sigma_{k}=u^{0} I+u^{a}\, \Sigma_{a}$, ($a=1,2, \cdots, n-1$) is
the following system of equations:
\begin{equation}\label{M_NMKdV}
u^{i}_{t}=u^{i}_{xxx}+6f_{jk}^{\ell}f_{r\ell}^iu^ju^ku_x^r,~~(i=0,1,2, \cdots, n-1),
\end{equation}
and the components of the recursion operator are given as
\begin{equation}
{\cal{R}}^{a}_{b}=\delta^{a}_{b} D^2+4 f^{a}_{bk}\, f^{k}_{\ell m}\, u^{\ell} u^{m}\,+4 f^{a}_{bk}\, f^{k}_{\ell m}\, u_x^{\ell}\, D^{-1}\, u^{m} ,~~~(a,b=0,1,\cdots, n-1),
\end{equation}
which gives consistently
\begin{equation}
u^{a}_{t}={\cal{R}}^{a}_{b}\, u^{b}_{x},  ~~~(a=0,1,\cdots,n-1).
\end{equation}

\noindent \textbf{Case n=2.} Consider the ${\cal M}_{2}$-extension of the MKdV equation. Let $u\rightarrow U=uI+v\Sigma_1$. We have $U_t=U_{xxx}+6U^2U_x$ which corresponds to the following coupled MKdV system of two equations:
\begin{align}
&u_t=u_{xxx}+6u^2u_x+12\beta uvv_x+6\alpha\beta v^2v_x+6\beta v^2u_x,\\
&v_t=v_{xxx}+6(u^2v)_x+6\alpha (v^2u)_x+6(\alpha^2+\beta)v^2v_x.
\end{align}
Let $v^2+v_xD^{-1}v=V_1$ and $v_xD^{-1}u+u_xD^{-1}v=V_2$. Then the recursion operator of the above system is
\begin{equation}
{\cal{R}}=\left(\begin{array}{ll}
        R_{MKdV}+4\beta V_1 &\quad 4\beta[2uv+\alpha V_1+V_2]   \cr
        4[2uv+\alpha V_1+V_2] &\quad R_{MKdV}+4[2\alpha uv+(\beta+\alpha^2)V_1+4\alpha V_2]
                \end{array} \right).
\end{equation}

\vspace{0.5cm}
\noindent \textbf{Case n=3.} Let us find the ${\cal M}_{3}$-extension of the MKdV equation (\ref{MKdV}). Let $u\rightarrow U=uI+v\Sigma_1+w\Sigma_2$. The equation $U_t=U_{xxx}+6U^2U_x$ gives the coupled MKdV system
of three equations as
{\small \begin{align}
&u_t=u_{xxx}+6u^2u_x-12\beta (uvw)_x-6\alpha\beta(v^2w)_x+6\beta(\gamma-\alpha^2)(w^2v)_x-6\alpha\beta(w^2u)_x\nonumber\\
&+6\beta(\beta+2\alpha\gamma-\alpha^3)w^2w_x-6\beta v^2v_x,\\
&v_t=v_{xxx}+6(u^2v)_x-12\gamma(uvw)_x-6(\beta+\alpha\gamma)(w^2u)_x+6(\gamma^2-\alpha\beta-\alpha^2\gamma)(w^2v)_x\nonumber\\
&-6(\beta+\alpha\gamma)(v^2w)_x+6(\beta+\alpha\gamma)(2\gamma-\alpha^2)w^2w_x-6\gamma v^2v_x,\\
&w_t=w_{xxx}+6(u^2w)_x+6(v^2u)_x+12\alpha (uvw)_x+6(\alpha^2-\gamma)(v^2w)_x+6(\alpha^3-2\alpha\gamma-\beta)(w^2v)_x\nonumber\\
&+6(\alpha^2-\gamma)(w^2u)_x+6(\gamma^2+\alpha^4-3\alpha^2\gamma-2\alpha\beta)w^2w_x+6\alpha v^2v_x.
\end{align}}
Recursion operator of the above system is
\begin{equation}
{\cal{R}}=\left(\begin{array}{lll}
        K_{11}&\quad K_{12} &\quad K_{13}  \cr
        K_{21} &\quad K_{22} &\quad K_{23}\cr
        K_{31}&\quad K_{32} &\quad K_{33}
                \end{array} \right),
\end{equation}
where{\small
\begin{align}
&K_{11}=R_{MKdV}+4B_{11}-4\beta[w_xC_{21}+(v_x+\alpha w_x)C_{31}],\\
&K_{12}=4B_{12}+4[u_xC_{12}-\beta w_x(D^{-1}u+C_{22})-\beta(v_x+\alpha w_x)C_{32}],\\
&K_{13}=4B_{13}+4[u_xC_{13}-\beta w_xC_{23}-\beta(v_x+\alpha w_x)(D^{-1}u+C_{33})],
\end{align}}
{\small
\begin{align}
&K_{21}=4B_{21}+4[v_xD^{-1}u-(u_x-\gamma w_x)C_{21}-(\gamma v_x+(\alpha\gamma+\beta) w_x)C_{31}],\\
&K_{22}=R_{MKdV}+4B_{22}+4[v_xC_{12}+(u_x-\gamma w_x)C_{22}-\gamma w_xD^{-1}u-(\gamma v_x+(\alpha \gamma+\beta)w_x)C_{32}],\\
&K_{23}=4B_{23}+4[v_xC_{13}+(u_x-\gamma w_x)C_{23}-(\gamma v_x+(\alpha\gamma+\beta) w_x)(D^{-1}u+C_{33})],\\
&K_{31}=4B_{31}+4[w_xD^{-1}u+(v_x+\alpha w_x)C_{21}+(u_x+\alpha v_x+(\alpha^2-\gamma) w_x)C_{31}],\\
&K_{32}=4B_{32}+4[w_xC_{12}+(v_x+\alpha w_x)(D^{-1}u+C_{22})+(u_x+\alpha v_x+(\alpha^2-\gamma) w_x)C_{32}],\\
&K_{33}=R_{MKdV}+4B_{33}+4[w_xC_{13}+(v_x+\alpha w_x)C_{23}+u_xC_{33}+(\alpha v_x+(\alpha^2-\gamma) w_x)(D^{-1}u+C_{33})],
\end{align}}
for
{\small
\begin{align}
&B_{11}=-2\beta wv-\alpha\beta w^2,\\
&B_{12}=\beta(\gamma-\alpha^2)w^2-2\beta(u+\alpha v)w-\beta v^2,\\
&B_{13}=\beta(\beta+2\alpha \gamma-\alpha^3)w^2+2\beta[(\gamma-\alpha^2)v-\alpha u]w-\alpha\beta v^2-2\beta uv,\\
&B_{21}=-(\beta+\alpha\gamma)w^2-2\gamma vw+2uv,\\
&B_{22}=(\gamma^2-\alpha^2\gamma-\alpha\beta)w^2-2[(\beta+\alpha\gamma)v+u\gamma]w-\gamma v^2,\\
&B_{23}=(2\gamma-\alpha^2)(\beta+\alpha\gamma)w^2-2[(\beta\alpha-\gamma^2+\alpha^2\gamma)v+(\beta+\alpha\gamma)u]w-(\beta+\alpha\gamma)v^2-2\gamma uv,\\
&B_{31}=(\alpha^2-\gamma)w^2+2(u+\alpha v)w+v^2,\\
&B_{32}=(\alpha^3-2\alpha\gamma-\beta)w^2-2[(\gamma-\alpha^2)v-\alpha u]w+2uv+\alpha v^2,\\
&B_{33}=(\alpha^4+\gamma^2-2\alpha\beta-3\alpha^2\gamma)w^2+2[(\alpha^3-\beta-2\alpha\gamma)v+(\alpha^2-\gamma)u]w+(\alpha^2-\gamma)v^2+2\alpha uv,
\end{align}}
and{\small
\begin{align}
&C_{12}=-\beta D^{-1}w,\quad C_{13}=-\beta D^{-1}(v+\alpha w),\quad C_{21}=D^{-1}v,\quad C_{22}=D^{-1}(u-\gamma w),\\
&C_{23}=-\gamma D^{-1}v-(\alpha\gamma+\beta)D^{-1}w,\quad C_{31}=D^{-1}w,\quad C_{32}=D^{-1}(v+\alpha w),\\
&C_{33}=D^{-1}u+\alpha D^{-1}v+(\alpha^2-\gamma)D^{-1}w.
\end{align}}

We have also studied SK and KK equations, and NLS system, their ${\cal M}_{n}$-extensions and recursion operators.
Since the expressions are quite longer we shall not give the recursion operators of the extensions of SK, KK, and NLS equations for $n=2, 3$ explicitly.

\section{SK system}

 The SK equation and its recursion operator are given as
\begin{eqnarray}
&& u_{t}+u_{5x}+5 u u_{xxx}+5 u_{x} u_{xx}+5 u^2 u_{x}=0, \label{SKeqn} \\
&& R_{SK}=D^6+6u D^4+9 u_{x} D^3+(9 u^2+11 u_{xx}) D^2+(10 u_{xxx}+21 u u_{x}) D\nonumber\\
&&\hspace{0.8cm}+4 u^3+16 u u_{xx}+6u_{x}^2+5 u_{4x}+u_{x} D^{-1}\,(2 u_{xx}+u^2)-u_{t}\, D^{-1}.
\end{eqnarray}

The ${\cal M}_{n}$-extension of the SK equation \cite{gur-pek2} is
the following system of equations:
\begin{equation}\label{M_NSK}
u^{i}_{t}=F^{i}\equiv -(u^{i}_{5x}+5f_{jk}^iu^ju_x^k+5f_{jk}^iu_x^ju_{xx}^k+5f_{jk}^{\ell}f_{r\ell}^iu^ju^ku_x^r),~~(i=0,1,2, \cdots, n-1).
\end{equation}
The components of the recursion operator of the above system are given as
\begin{align}
&{\cal{R}}^{a}_{b}=\delta^{a}_{b} D^6+6 f^{a}_{bc}u^cD^4+9 f^{a}_{bc} u_x^{c}D^3+(9f_{bk}^af_{\ell m}^ku^{\ell}u^m+11f_{bc}^au_{xx}^c)D^2
\nonumber\\
&+(10f_{bc}^au_{xxx}^c+21f_{bk}^af_{\ell m}^ku^{\ell}u_x^m)D+4f_{bk}^af_{\ell m}^kf_{rs}^mu^{\ell}u^ru^s
+16f_{bk}^af_{\ell m}^ku^{\ell}u_{xx}^m+6f_{bk}^af_{\ell m}^ku_x^{\ell}u_x^m\nonumber\\
&+5f_{bc}^au_{4x}^c+f_{bc}^au_x^cD^{-1}(2f_{\ell m}^ku_{xx}^m+f_{ij}^df_{rs}^ju^ju^r)-f_{bc}^a\, F^c\,D^{-1},
\end{align}
for $(a,b=0,1,\cdots, n-1)$ producing higher symmetries consistently
\begin{equation}
u^{a}_{t}={\cal{R}}^{a}_{b}\, u^{b}_{x},  ~~~(a=0,1,\cdots,n-1).
\end{equation}
The SK system is obtained by using the trivial symmetry $u^{a}_{t}={\cal{R}}^{a}_{b}\,(0)^{b}$.
\medskip

\noindent \textbf{Case n=2}. Letting $u\rightarrow U=u\Sigma_0+v\Sigma_1$ where $\Sigma_0=I$ and $\Sigma_1$ is given by (\ref{Sigma_1n=2}) the SK equation (\ref{SKeqn}) becomes
\begin{align}
&u_t+u_{5x}+5(uu_{xx}+\beta vv_{xx}+\beta uv^2)_x+5u^2u_x+5\alpha\beta v^2 v_x=0,\\
&v_t+v_{5x}+5(uv_{xx}+vu_{xx}+\alpha vv_{xx}+u^2v+\alpha v^2 u)_x+5(\alpha^2+\beta)v^2 v_x=0.
\end{align}

\noindent \textbf{Case n=3}. The ${\cal M}_{3}$-extension of SK equation is derived by letting $u\rightarrow U=uI+v\Sigma_1+w\Sigma_2$ where $\Sigma_1$ is given by (\ref{Sigma_1n=3}) and $\Sigma_2$ is as in (\ref{Sigma_2n=3}). The SK equation (\ref{SKeqn}) becomes
{\small\begin{align}
&u_t+u_{5x}+5u^2u_x+5(uu_{xx})_x+5\beta(\gamma-\alpha^2)(w^2v)_x-5\alpha\beta(w^2u)_x-5\alpha\beta(v^2w)_x-5\beta v^2v_x\nonumber\\
&+5\beta(\beta-\alpha^3+2\alpha\gamma)w^2w_x-5\alpha\beta(ww_{xx})_x-5\beta(vw_{xx})_x-5\beta(wv_{xx})_x-10\beta(uvw)_x=0,\\
&v_t+v_{5x}+5(uv_{xx})_x+5(vu_{xx})_x+5(u^2v)_x-5(\beta+\alpha\gamma)(w^2u)_x-5(\beta+\alpha\gamma)(v^2w)_x-5\gamma v^2v_x\nonumber\\
&+5(\gamma^2-\alpha^2\gamma-\alpha\beta)(w^2v)_x-5(\alpha^2-2\gamma)(\alpha\gamma+\beta)w^2w_x-5(\beta+\alpha\gamma)(ww_{xx})_x\nonumber\\
&-5\gamma(wv_{xx})_x-5\gamma(vw_{xx})_x-10\gamma(uvw)_x=0,\\
&w_t+w_{5x}+5(uw_{xx})_x+5(vv_{xx})_x+5(wu_{xx})_x+5(u^2w)_x+5(v^2u)_x+5(\alpha^2-\beta-2\alpha\gamma)(w^2v)_x\nonumber\\
&+5(\alpha^2-\gamma)(w^2u)_x+5(\alpha^2-\gamma)(v^2w)_x+5(\alpha^4+\gamma^2-\alpha^2\gamma-2\alpha\beta)w^2w_x+5(\alpha^2-\gamma)(ww_{xx})_x\nonumber\\
&+5\alpha(vw_{xx})_x+5\alpha(wv_{xx})_x+5\alpha v^2v_x+10\alpha(uvw)_x=0.
\end{align}}

\section{KK system}

The KK equation and its recursion operator are given as
\begin{eqnarray}
&& u_{t}+u_{5x}+10 u u_{xxx}+25 u_{x} u_{xx}+20 u^2 u_{x}=0, \label{KKeqn}\\
&& R_{KK}=D^6+12 u D^4+36 u_{x} D^3+(36 u^2+49 u_{xx}) D^2+(35 u_{xxx}+120 u u_{x}) D\nonumber\\
&&\hspace{0.8cm}+32 u^3+82 u u_{xx}+69 u_{x}^2+13 u_{4x}+2u_{x} D^{-1}\,( u_{xx}+4 u^2)-2u_{t}\, D^{-1}.
\end{eqnarray}

The ${\cal M}_{n}$-extension of the KK equation \cite{gur-pek2} is represented by
\begin{equation}\label{M_NKK}
u^{i}_{t}=G^{i}\equiv -(u^{i}_{5x}+10f_{jk}^iu^ju_{xxx}^k+25f_{jk}^iu_x^ju_{xx}^k+20f_{jk}^{\ell}f_{r\ell}^iu^ju^ku_x^r),~~(i=0,1,2,\cdots,n-1).
\end{equation}
The components of the recursion operator of the above system are
\begin{align}
&{\cal{R}}^{a}_{b}=\delta^{a}_{b} D^6+12 f^{a}_{bc}u^cD^4+36 f^{a}_{bc} u_x^{c}D^3+(36f_{bk}^af_{\ell m}^ku^{\ell}u^m+49f_{bc}^au_{xx}^c)D^2\nonumber\\
&+(35f_{bc}^au_{xxx}^c+120f_{bk}^af_{\ell m}^ku^{\ell}u_x^m)D+32f_{bk}^af_{\ell m}^kf_{rs}^mu^{\ell}u^ru^s+82f_{bk}^af_{\ell m}^ku^{\ell}u_{xx}^m
\nonumber\\
&+69f_{bk}^af_{\ell m}^ku_x^{\ell}u_x^m+13f_{bc}^au_{4x}^c+2f_{bc}^au_x^cD^{-1}(f_{\ell m}^ku_{xx}^m+4f_{ij}^df_{rs}^ju^ju^r)-2f_{bc}^a\,G^c\,D^{-1},
\end{align}
for $(a,b=0,1,\cdots, n-1)$ giving higher symmetries consistently
\begin{equation}
u^{a}_{t}={\cal{R}}^{a}_{b}\, u^{b}_{x},  ~~~(a=0,1,\cdots,n-1).
\end{equation}
The KK system is obtained by using the trivial symmetry $u^{a}_{t}={\cal{R}}^{a}_{b}\,(0)^{b}$.

\medskip

\noindent \textbf{Case n=2.} The ${\cal M}_{2}$-extension of the KK equation can be obtained by taking $u\rightarrow U=u\Sigma_0+v\Sigma_1$ where $\Sigma_0=I$ and $\Sigma_1$ is given by (\ref{Sigma_1n=2}). We have
\begin{align}
&u_t+u_{5x}+10uu_{xxx}+25u_xu_{xx}+\beta(10vv_{xxx}+25v_xv_{xx})+20u^2u_x+20\beta (v^2u)_x+20\alpha\beta v^2v_x=0,\\
&v_t+v_{5x}+10uv_{xxx}+25u_xv_{xx}+10vu_{xxx}+25v_xu_{xx}+\alpha(10vv_{xxx}+25v_xv_{xx})+20(u^2v)_x+20\alpha(v^2u)_x\nonumber\\
&+20(\alpha^2+\beta)v^2v_x=0.
\end{align}

\noindent \textbf{Case n=3}. Let us now derive the ${\cal M}_{3}$-extension of KK equation. Take $u\rightarrow U=uI+v\Sigma_1+w\Sigma_2$. The KK equation
(\ref{KKeqn}) turns to be
{\small \begin{align}
&u_t+u_{5x}+25u_xu_{xx}+10uu_{xxx}+20u^2u_x+20\beta(\beta-\alpha^3+2\alpha\gamma)w^2w_x-20\beta v^2v_x+20\beta(\gamma-\alpha^2)(w^2v)_x\nonumber\\
&-20\alpha\beta(w^2u)_x-20\alpha\beta(v^2w)_x-25\beta(w_xv_x)_x-10\alpha\beta ww_{xx}-25\alpha\beta w_xw_{xx}-10\beta wv_{xxx}\nonumber\\
&-10\beta vw_{xx}-40\beta(uvw)_x=0,\\
&v_t+v_{5x}+10uv_{xxx}+10vu_{xxx}+25(u_xv_x)_x+20(u^2v)_x-25(\beta+\alpha\gamma)w_xw_{xx}-20(\beta+\alpha\gamma)(w^2u)_x\nonumber\\
&-20(\alpha^2-2\gamma)(\beta+\alpha\gamma)w^2w_x+20(\gamma^2-\alpha^2\gamma-\alpha\beta)(w^2v)_x-20(\beta+\alpha\gamma)(v^2w)_x-10(\beta+\alpha\gamma)ww_{xxx}
\nonumber\\
&-10\gamma vw_{xxx}-10\gamma wv_{xxx}-25\gamma v_xw_{xx}-25\gamma w_xv_{xx}-20\gamma v^2v_x-40\gamma(uvw)_x=0,\\
&w_t+w_{5x}+10uw_{xxx}+10vv_{xxx}+10wu_{xxx}+25u_xw_{xx}+25v_xv_{xx}+25w_xu_{xx}+20(u^2w)_x+20(v^2u)_x
\nonumber\\&+10(\alpha^2-\gamma)ww_{xxx}+25(\alpha^2-\gamma)w_xw_{xx}+20(\alpha^4+\gamma^2-3\alpha^2\gamma-2\alpha\beta)w^2w_x+20(\alpha^2-\gamma)(w^2u)_x\nonumber\\
&-20(2\alpha\gamma-\alpha^3+\beta)(w^2v)_x+20(\alpha^2-\gamma)(v^2w)_x+10\alpha vw_{xxx}+10\alpha wv_{xxx}\nonumber\\
&+25\alpha (v_xw_x)_x+40\alpha(uvw)_x=0.
\end{align}}

\section{NLS system}

So far we have applied our method to scalar equations. Here we give an example where we can also apply it to a system of equations. NLS system is given as follows \cite{AKNS}-\cite{gur-pek4}
\begin{eqnarray}
&&a q_{t}=\frac{1}{2} q_{xx}-q^2 r,\label{NLS1} \\
&&a r_{t}=-\frac{1}{2} r_{xx}+r^2 q,\label{NLS2}
\end{eqnarray}
with recursion operator
\begin{equation}
R_{NLS}=\left( \begin{array}{cc}
\frac{1}{2}\,D-q D^{-1}\,r&-q D^{-1} q \\
r D^{-1} r&-\frac{1}{2}\,D+r D^{-1}\,q\\
\end{array} \right).
\end{equation}
\medskip
Letting $q \to Q=q^{0} I+q^{\alpha} \Sigma_{\alpha}$ and $r \to R=r^{0} I+r^{\alpha} \Sigma_{\alpha}$, ($\alpha=1,2, \cdots, n-1$), we obtain a system of NLS equations,
\begin{eqnarray}
&&a q^{i}_{t}=\frac{1}{2}\, q^{i}_{xx}-f^{i}_{mj}\,f^{m}_{k \ell}\, q^{k}\, q^{\ell}\, r^{j}, ~~(i=0,1, \cdots, n-1),\\
&&a r^{i}_{t}=-\frac{1}{2}\, q^{i}_{xx}+f^{i}_{mj}\,f^{m}_{k \ell}\, r^{k}\, r^{\ell}\, q^{j}, ~~(i=0,1, \cdots, n-1),
\end{eqnarray}
where the components of the recursion operator of the above system are given as follows:
\begin{equation}
a U^{A}_{t}={\cal{R}}^{A}_{B}\, U^{B}_{x},
\end{equation}
for $U^{A}=(q^{i}, r^{\alpha})$, ($i, \alpha=0,1,\cdots,n-1$), and
\begin{equation}
{\cal{R}}^{A}_{B}=\left( \begin{array}{cc}
{\cal{R}}^{i}_{j}& {\cal{R}}^{i}_{\alpha}\\
{\cal{R}}^{\beta}_{j}&{\cal{R}}^{\beta}_{\alpha}\\
\end{array} \right),\quad (i,j,\alpha, \beta=0,1,\cdots, n-1),
\end{equation}
with
\begin{eqnarray}
&&{\cal{R}}^{i}_{j}=\frac{1}{2}\,D\, \delta^{i}_{j}-f^{i}_{jk}\,f^{k}_{m \alpha}\, q^{m}\, D^{-1}\, r^{\alpha}, \\
&&{\cal{R}}^{i}_{\alpha}=-f^{i}_{\alpha k}\,f^{k}_{mn}\, q^m D^{-1} q^n, \\
&&{\cal{R}}^{\beta}_{j}=f^{\beta}_{jk}\,f^{k}_{\alpha \gamma}\, r^\alpha D^{-1} r^\gamma, \\
&&{\cal{R}}^{\beta}_{\alpha}=-\frac{1}{2}\,D\, \delta^{\beta}_{\alpha}+f^{\beta}_{\alpha k}\,f^{k}_{\gamma n}\, r^{\gamma}\, D^{-1}\, q^{n}.
\end{eqnarray}


\noindent
The NLS systems corresponding to $n=2,3$ are given as:

\noindent \textbf{Case n=2}. Let us derive ${\cal M}_{2}$-extension of the NLS system (\ref{NLS1})-(\ref{NLS2}). Take $q \to Q=uI+v\Sigma_1$ and $r \to R=w I+s \Sigma_1$
where $\Sigma_1$ is given in (\ref{Sigma_1n=2}). Then we obtain the following NLS system of four equations:
\begin{align}\displaystyle
&au_t=\frac{1}{2}u_{xx}-u^2w-2\beta uvs-\alpha\beta v^2s-\beta v^2w,\\
&av_t=\frac{1}{2}v_{xx}-u^2s-2uvw-2\alpha uvs-\alpha v^2w-(\alpha^2+\beta)v^2s,\\
&aw_t=-\frac{1}{2}w_{xx}+w^2u+2\beta wsv+\alpha \beta s^2v+\beta s^2 u,\\
&as_t=-\frac{1}{2}s_{xx}+w^2v+2wsu+2\alpha wsv+\alpha s^2u+(\alpha^2+\beta)s^2v.
\end{align}

\noindent \textbf{Case n=3}. Consider now ${\cal M}_{3}$-extension of the NLS system (\ref{NLS1})-(\ref{NLS2}). Letting $q \to Q=uI+v\Sigma_1+z\Sigma_2$ and $r \to R=w I+s \Sigma_1+p\Sigma_2$ we have the NLS system of six equations:
{\small \begin{align}\displaystyle
&au_t=\frac{1}{2}u_{xx}-u^2w+\beta(\alpha^2-\gamma)z^2s+\beta\alpha(\alpha^2-\gamma)z^2p+\alpha\beta v^2p-\beta^2z^2p+\beta v^2s+\alpha\beta z^2w\nonumber\\
&+2\beta(\alpha^2-\gamma)vzp+2\alpha\beta vzs+2\alpha\beta uzp+2\beta uzs+2\beta uvp+2\beta vzw,\\
&av_t=\frac{1}{2}v_{xx}-u^2s-2uvw+(\alpha\beta+\alpha^2\gamma-\gamma^2)z^2s+(\beta+\alpha\gamma)z^2w+(\beta+\alpha\gamma)v^2p+\gamma v^2s\nonumber\\
&+(\beta+\alpha\gamma)(\alpha^2-2\gamma)z^2p+2(\beta+\alpha\gamma)uzp+2(\alpha\beta-\gamma^2+\alpha^2\gamma)vzp+2(\beta+\alpha\gamma)vzs\nonumber\\
&+2\gamma uvp+2\gamma vzw+2\gamma uzs,\\
&az_t=\frac{1}{2}z_{xx}-u^2p-v^2w-2uvs-2uzw+(\gamma-\alpha^2)v^2p+(\gamma-\alpha^2)z^2w+(p-\alpha^3+2\alpha\gamma)z^2s\nonumber\\
&+(3\alpha^2\gamma-\alpha^4+2\alpha\beta-\gamma^2)z^2p-\alpha v^2s+2(\gamma-\alpha^2)uzp+2(\gamma-\alpha^2)vzs\nonumber\\
&+2(2\alpha\gamma-\alpha^3+\beta)vzp-2\alpha uzs-2\alpha uvp-2\alpha vzw,
\end{align}}
\small{
\begin{align}
&aw_t=-\frac{1}{2}w_{xx}+w^2u+(2\alpha\beta\gamma+\beta^2-\alpha^3\beta)p^2z+\beta(\gamma-\alpha^2)p^2v-\beta s^2v-\alpha\beta s^2z-\alpha\beta p^2u\nonumber\\
&+2\beta(\gamma-\alpha^2)spz-2\beta wsz-2\beta spu-2\beta wpv-2\alpha\beta wpz-2\alpha\beta spv,\\
&as_t=-\frac{1}{2}s_{xx}+w^2v+2wsu+(\alpha\gamma+\beta)(2\gamma-\alpha^2)p^2z+(\gamma^2-\alpha\beta-\alpha^2\gamma)p^2v-(\alpha\gamma+\beta)s^2z\nonumber\\
&-(\alpha\gamma+\beta)p^2u-\gamma s^2v-2(\alpha\gamma+\beta)wpz+2(\gamma^2-\alpha\beta-\alpha^2\gamma)spz-2(\alpha\gamma+\beta)spv\nonumber\\
&-2\gamma spu-2\gamma wpv-2\gamma wsz,\\
&ap_t=-\frac{1}{2}p_{xx}+s^2u+w^2z+2wsz+2wpu+(\alpha^4+\gamma^2-3\alpha^2\gamma-2\alpha\beta)p^2z+(\alpha^3-\beta-2\alpha\gamma)p^2v\nonumber\\
&+(\alpha^2-\gamma)p^2u+(\alpha^2-\gamma)s^2z+\alpha s^2v+2(\alpha^2-\gamma)wpz+2(\alpha^2-\gamma)spv+2(\alpha^3-\beta-2\alpha\gamma)spz\nonumber\\
&+2\alpha wsz+2\alpha spu+2\alpha wpv.
\end{align}}

\section{Concluding remarks}
We continue to apply our ${\cal M}_{n}$-extension method by using Frobenius companion matrices whose powers form a closed commutative
algebra under the matrix multiplication. We derived new integrable systems and their recursion operators from integrable scalar equations. These new systems  are written in terms of structure constants of this algebra. The structure constants are indeed the components of companion matrices. We obtained the structure constants for $n=2,3,4$.
For these cases, we gave the members of basis for the algebra constructed by the powers of companion matrices, explicitly. By using our method with the algebras for $n=2,3$ we presented KdV, MKdV, SK, KK, and NLS systems. Most of these systems are new. It will be highly interesting to study the local and nonlocal reductions of these systems.\\

\end{document}